\documentclass[reprint, letter, groupedaddress]{revtex4-1}

\usepackage{color}
\usepackage{graphicx}
\usepackage{amsmath}
\usepackage{bm}
\usepackage{float}
\usepackage{upgreek}
\usepackage{xspace}
\usepackage{hyperref}

\newcommand{\micron}{$\upmu$m\xspace}
\newcommand{\microJ}{$\upmu$J\xspace}
\newcommand{\etal}{et~al.}

\newcommand{\revision}[1]{#1}

\begin{document}

\title{Generation of heralded single photons beyond 1100~nm by spontaneous four-wave mixing
in a side-stressed femtosecond laser-written waveguide }

\author{Zhizhong Yan}
\email{zhizhong.yan@mq.edu.au}
\author{Yuwen Duan}
    \thanks{Currently with the State Key Laboratory of Information Photonics and Optical Communications, Beijing University of Posts and Telecommunications (BUPT); and the School of Ethnic Minority Education, BUPT, 10 Xitucheng Road, Beijing, China 100876.}
\author{L.~G. Helt}
\author{Martin Ams}
\author{Michael J. Withford}
\author{M.~J. Steel}
\email{michael.steel@mq.edu.au}
\affiliation{Centre for Ultrahigh bandwidth Devices for Optical Systems (CUDOS), MQ Photonics
Research Centre, Department of Physics and Astronomy, Macquarie University, NSW 2109, Australia}

\date{\today}

\begin{abstract} We demonstrate a monolithically integrable heralded photon source in a femtosecond laser direct written glass waveguide. The generation of photon pairs with a
wide wavelength separation requires a concomitant large birefringence in the
normal dispersion regime. Here, by incorporation of side-stress tracks,  we produce
a waveguide with a birefringence of $1.64\times~10^{-4}$ and propagation
loss as low as 0.21 dB/cm near 980~nm. We measure photon pairs with 300~nm wavelength separation at an internal generation rate exceeding $5.05\times10^6$/s. The second order correlations indicate that the generated photon pairs are in a strongly non-classical regime.
\end{abstract}

\maketitle

As the field of quantum integrated photonics has emerged over the last decade,
a range of device platforms have been explored. One approach that offers
complementary capabilities to traditional planar technologies is the direct
inscription of waveguides in glass by a femtosecond
laser~\cite{Davis1996,Gattass2008,Ams2009,Ams2012}. The ability of the femtosecond laser
direct write (FLDW) platform to write 3D circuits, manipulate polarization and
incorporate tunability has enabled demonstrations of numerous quantum
phenomena~\cite{Meany2015} including multiple photon
interference~\cite{Marshall2009}, quantum walks~\cite{Sansoni2010a,Owens2011}, boson
sampling~\cite{Tillmann2013,Crespi2013a}, complex state
preparation~\cite{Grafe2014}, and novel
interferometry~\cite{Meany2012,Spagnolo2012,Chaboyer2015}.

While the FLDW approach has proven flexible for photon processing, ultimately
quantum photonics applications demand integration of photon sources and
detectors in the same device. A common approach in other chip platforms is to
use spontaneous nonlinear processes to generate correlated photon
pairs~\cite{Clark2016}. Detection of a herald photon in one wavelength band or
polarization state flags the imminent arrival of a partner photon in a
complementary band or mode. Given the ubiquity of telecom fibers in the
infrared and photon detectors in the visible, the two bands of 1530-1565~nm and
600-850~nm are both important target wavelength ranges for photon generation.
Indeed creating one photon in each band is attractive as the short wavelength
``signal'' may be detected with efficient silicon avalanche photodiodes (APD)
while the long wavelength ``idler'' is suitable for transport over fiber.  In
FLDW waveguides, however, the limited nonlinearity of silica-based glasses
($\sim$10$^{-20}~\text{m}^{2}/\text{W}$) and typical waveguide lengths (under
$\sim$10~cm), mean a photon generation strategy based on nonlinear optics might
seem challenging.  Nonetheless, classical nonlinear optics in laser written
waveguides has been explored to a limited extent~\cite{Blomer2006, Psaila2007}.
Moreover, some fibre sources of photon pairs are only centimeters
long~\cite{Fang2013,Fang2014} and have optical mode areas comparable to FLDW
waveguides~\cite{McMillan2013}, suggesting that correlated pair generation in
silica FLDW waveguides is in fact viable.

In 2013, Spring~\etal~\cite{Spring2013} reported photon pair generation by spontaneous
four wave mixing (SFWM) in a silica laser written waveguide. Their experiment
used a Ti:sapphire oscillator at $\lambda_p=729$~nm as a pump source to
generate signal and idler photons centered at $\lambda_s=676$~nm and
$\lambda_i=790$~nm. The device has the added benefit of avoiding contamination
from noise due to spontaneous Raman scattering which extends only up to $\sim$20~THz
from the pump, whereas the pump-idler separation was $\sim$64~THz.
A key to this experiment was the use of the form birefringence of the mildly elliptical waveguide to enable phase-matching despite the significant signal-idler separation.

In the present work, we take the next steps toward a telecom-band heralded
photon source in silica FLDW waveguides by generating widely-separated pairs
with the signal photons in the near-visible silicon APD band, and the idler
photons reaching the sensitivity range of InGaAsP single photon detectors.
Since the refractive index contrast of laser written waveguides is relatively
small~\cite{Marshall2009} (typically below $5\times 10^{-3}$), the potential to
significantly increase form birefringence so as to extend the signal-idler separation
is likely small. Furthermore, large
form birefringence also increases the coupling loss for monolithic integration
to quantum photonic circuits, which usually rely on waveguides with several
orders lower birefringence~\cite{Tillmann2013}.  Thus, to achieve the required
birefringence and provide a path towards even-larger birefringence and larger
frequency shifts, we exploit the concept of stress-induced birefringence
induced by a pair of additional FLDW tracks written parallel to the central
waveguide~\cite{Fernandes2012}. With a pump at $\lambda_p=957$~nm we generate
photons centered at $\lambda_s=830$~nm and $\lambda_i=1130$~nm, the signal
photons having a purity of approximately 0.93.

\begin{figure}[h]
  \includegraphics[width=\columnwidth]{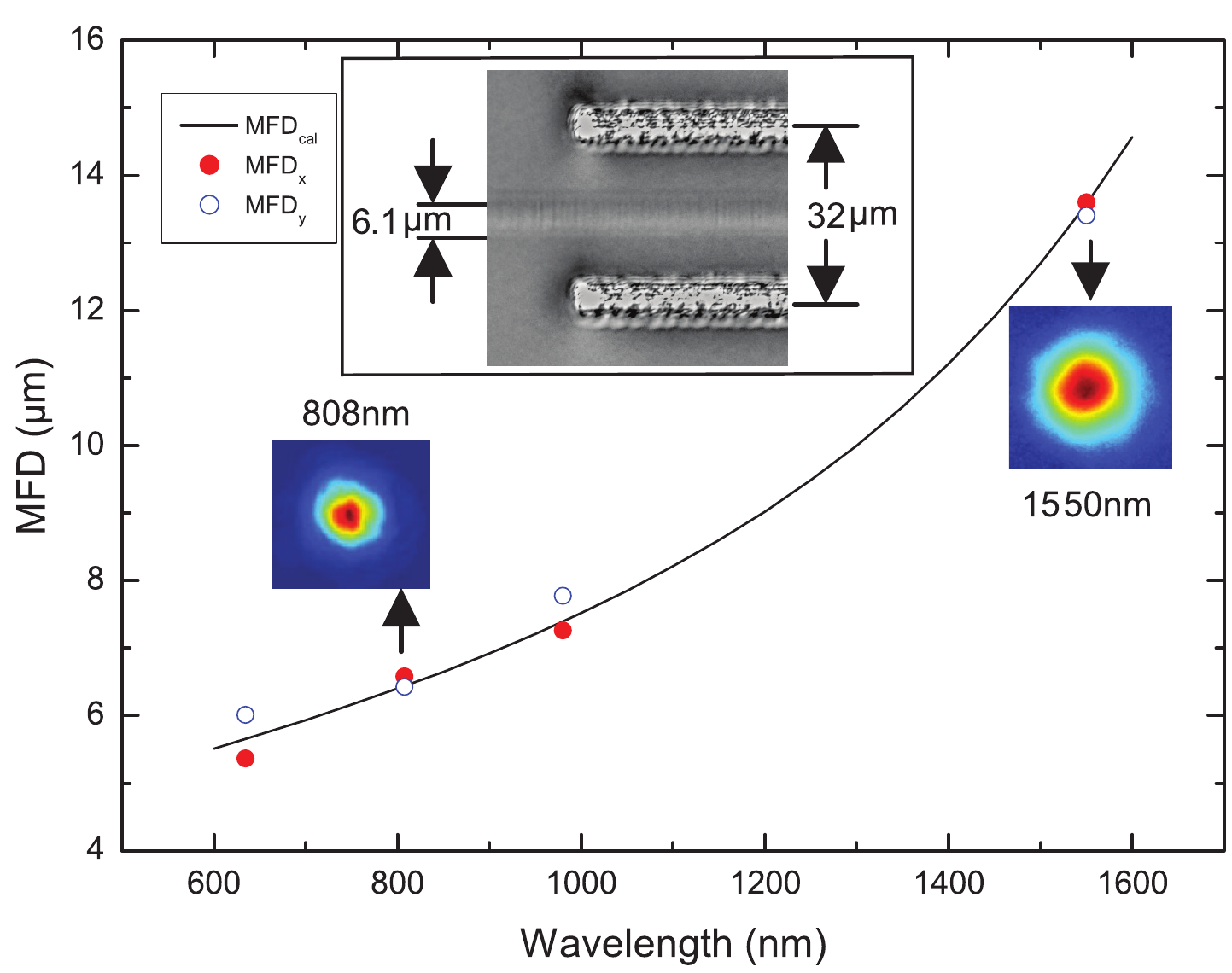}
  \caption{Measured MFD at four wavelengths (635~nm, 808~nm, 980~nm and 1550~nm) in
 horizontal (MFD$_x$) and vertical (MFD$_y$) directions. Mode profile images are shown
 for 808~nm and 1550~nm, respectively. The solid curve is a fit to the MFD
 based on a step-index approximation for the waveguide. The larger inset shows a
 differential interference contrast (DIC) microscopy image of the waveguide seen from above.
  }
  \label{fig:the_chip}
\end{figure}

The details of our device are as follows.
The central waveguide and side stress tracks (see  Fig.~\ref{fig:the_chip}
inset) were fabricated in a fused silica sample (Schott Lithosil) using a
regeneratively amplified Ti:sapphire laser at 800~nm (repetition rate 1 kHz,
pulse duration 120 fs). The laser beam was circularly polarized and focused
into the glass with a 20$\times$ (NA 0.46) microscope objective at a depth of
170~\micron below the surface.  A slit was placed in the beam path to produce
waveguides with a circular cross section~\cite{Ams2005}.  For the  central waveguides, eight
writing passes were used to achieve the lowest propagation losses, with a pulse
energy of 1.1~\microJ.  The total length of the central waveguide was 30~mm.
Similar writing conditions were used for the stress side tracks, except that no
slit was used and the pulse energy for the stress tracks was 0.3~\microJ, much
lower than that for writing the waveguides, while three writing passes were
applied.

The two stress tracks were positioned
16~$\mu$m either side of the central waveguide. The stress tracks begin and end
2~mm in from the input and output facets, so that the total effective length of
the region with enhanced birefringence is 26~mm.
The propagation loss was measured  by transmission measurements at three wavelengths,
subtracting estimates of the coupling loss obtained from mode profile measurements.
The measured losses were
0.29~dB/cm ($\lambda=808$~nm),
0.21~dB/cm (980~nm) and 2.9~dB/cm (1550~nm), where the much larger
loss at the longer wavelength is attributed to the more extended mode field
interacting strongly with the stress field.

As the precise stress field and consequent effects of the stress tracks is
difficult to model, we conducted a series of classical measurements to learn as much
as possible about the phase matching conditions in our device before proceeding to
quantum measurements. We first measured mode field diameters (MFDs) and
polarization rotations to obtain a rough estimate of the birefringence; then refined this
approximation with classical FWM measurements. Figure~\ref{fig:the_chip} shows
MFD measurements at four wavelengths.
The solid line is a fit to the
MFD as a function of $V$-number~\cite{marcuse1977loss} approximating the
waveguide as a step-index device.  Identifying an effective waveguide diameter
of 6.1~$\mu$m (consistent with the DIC microscope measurement as shown in the inset to
Fig.~\ref{fig:the_chip}), we extract an estimate of the peak refractive
index change in the central waveguide as $4\times 10^{-3}$.
%
%
\revision{Based on these measurements and silica material dispersion~\cite{web}, we can calculate the expected phase-matching curves as a function of the pump laser wavelength for several conceivable
values of birefringence, assuming that the birefringence is wavelength independent (see Fig.~\ref{fig:phasematch}). In fact, as seen below, there will likely be some reduction in the birefringence with wavelength, but the precise amount is not understood at this time.
}
The vertical line \revision{in Fig.~\ref{fig:phasematch}} indicates the phase-matched signal and idler wavelengths
if the pump is at 957~nm.

\begin{figure}[h]
\includegraphics[width=\columnwidth]{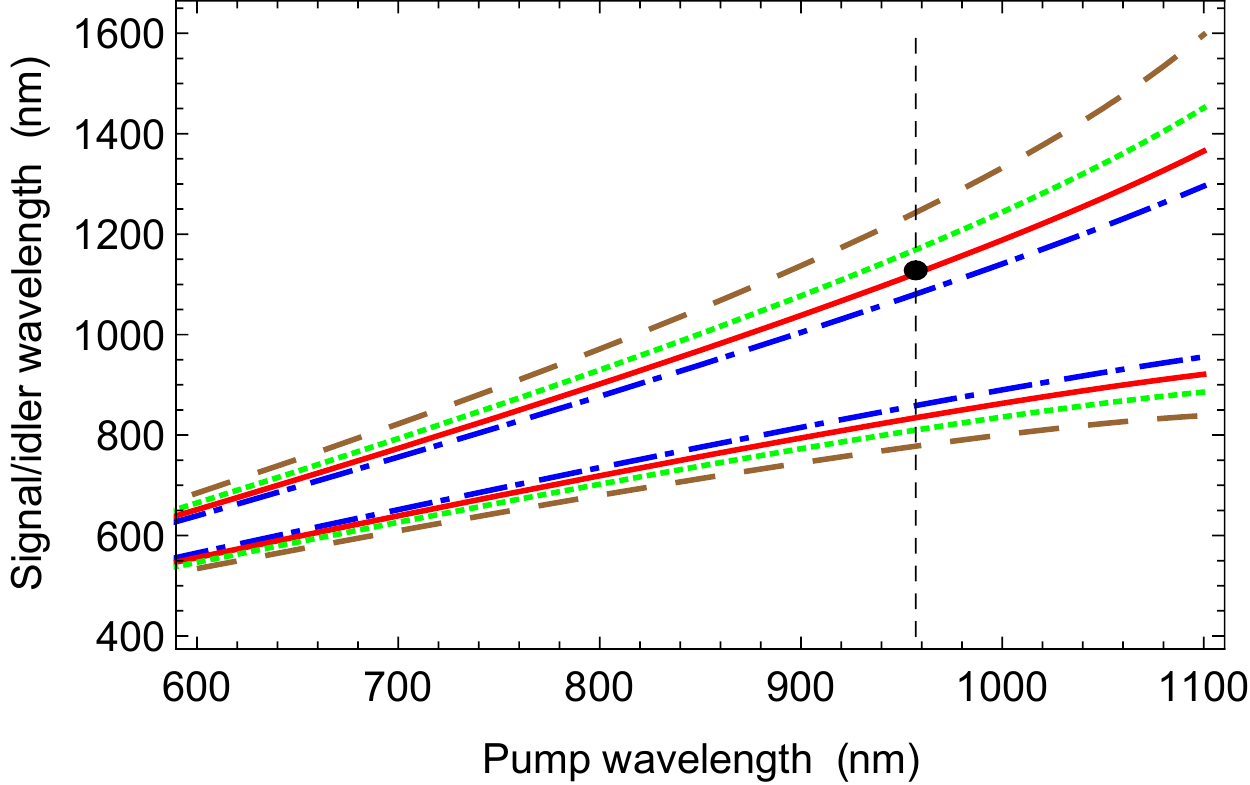}
\caption{
%
\revision{Calculated phase-matching curves under cross-polarisation conditions of a fused silica FLDW waveguide using the same dimensions in Fig.~\ref{fig:the_chip} assuming constant birefringence values of $1.0 \times 10^{-4}$ (blue, dash-dot line), $1.64 \times 10^{-4}$ (red, solid line), $2.5 \times 10^{-4}$ (green, dot line), and $4.0 \times 10^{-4}$ (brown, dash line).}
The dashed line and dots indicate the operating parameters for the quantum
characterization, with  the black dot marking the idler wavelength for a birefringence of $1.64\times 10^{-4}$.
} \label{fig:phasematch} \end{figure}

To obtain an estimate of the birefringence, as the expected beat length is on
the mm scale, we took polarization rotation measurements at a range of
wavelengths centered at 960~nm.  Simple analysis~\cite{Sansoni2010a} then
indicates a birefringence value of order $1.8\pm 0.3 \times
10^{-4}$ with the slow axis in the vertical direction (perpendicular to the
line joining the side-stress features).  To verify this measurement we
performed a classical four wave mixing experiment using a tunable femtosecond
Ti-Sapphire laser (Coherent Chameleon, 80 MHz repetition rate, 140~fs pulse
width) to pump the FLDW waveguide.
\revision{In the following, values of pump power were determined at
the output facet of the chip for
all parts of the experiment including the SFWM. We achieved more than 35~$\%$ throughput
from the input to the output facet at wavelengths near 800 nm with power readings over 500 mW
for the classical FWM.}
A continuous wave diode laser with wavelength
fixed at 976~nm was applied to seed the idler (long-wavelength) channel. The
pump was launched with vertical polarisation and the seed laser with horizontal
polarization with the input beams combined on a polarisation beam splitter
cube. The signal power spectrum at the output was collected by an optical
spectrum analyser (OSA) after polarisation filtering to remove the strong pump
power, followed by two short pass filters at 750~nm (cut-off suppression
$> 50$~dB each). Due to the short waveguide length, the phase matching condition
permitted a continuously observable signal spectrum above the noise level (-85
dBm) of the OSA over a pump tuning range of 838~nm to 848~nm. In the test, we
also switched off the pump and seed lasers separately to confirm the FWM
origin of the signal. By monitoring the power in  the signal band around
737~nm, we extracted a birefringence in a range from $1.9 \times 10^{-4}$ to
$2.13 \times 10^{-4}$, and verified the earlier identification of the
vertical axis as the slow axis.

In developing the correlated photon pair source, our goal was to push the idler
band as far as possible into the infrared entering the sensitivity regime of
InGaAs single photon detectors (SPDs). In this work, we used a free-running InGaAs SPD (id-Quantique id-220). At the same time we wished to keep the signal photons in the detection band of silicon
APD detectors. This was achieved by a shift in the pump wavelength into a range
around 950~nm, which raises the question of whether the level of birefringence
will be maintained with longer wavelength operation. Form birefringence, which
is dominated by the regions of most rapid index change, is commonly relatively
independent of wavelength. However, in our case the birefringence
derives from the spatial sampling by the optical mode of the stress field
around the central waveguide and the wavelength dependence of the mode area
(see Fig.~\ref{fig:the_chip}) is likely to vary this sampling and alter the
birefringence. Unfortunately, the longer pump wavelength leads to significantly
reduced available power ($< $150 mW) which was
insufficient to observe a classical FWM signal with the OSA.
Nevertheless, our classical characterization was sufficient to direct us to
the correct wavelength ranges for observing spontaneous four wave mixing.

\begin{figure}[htbp] \includegraphics[width=\columnwidth]{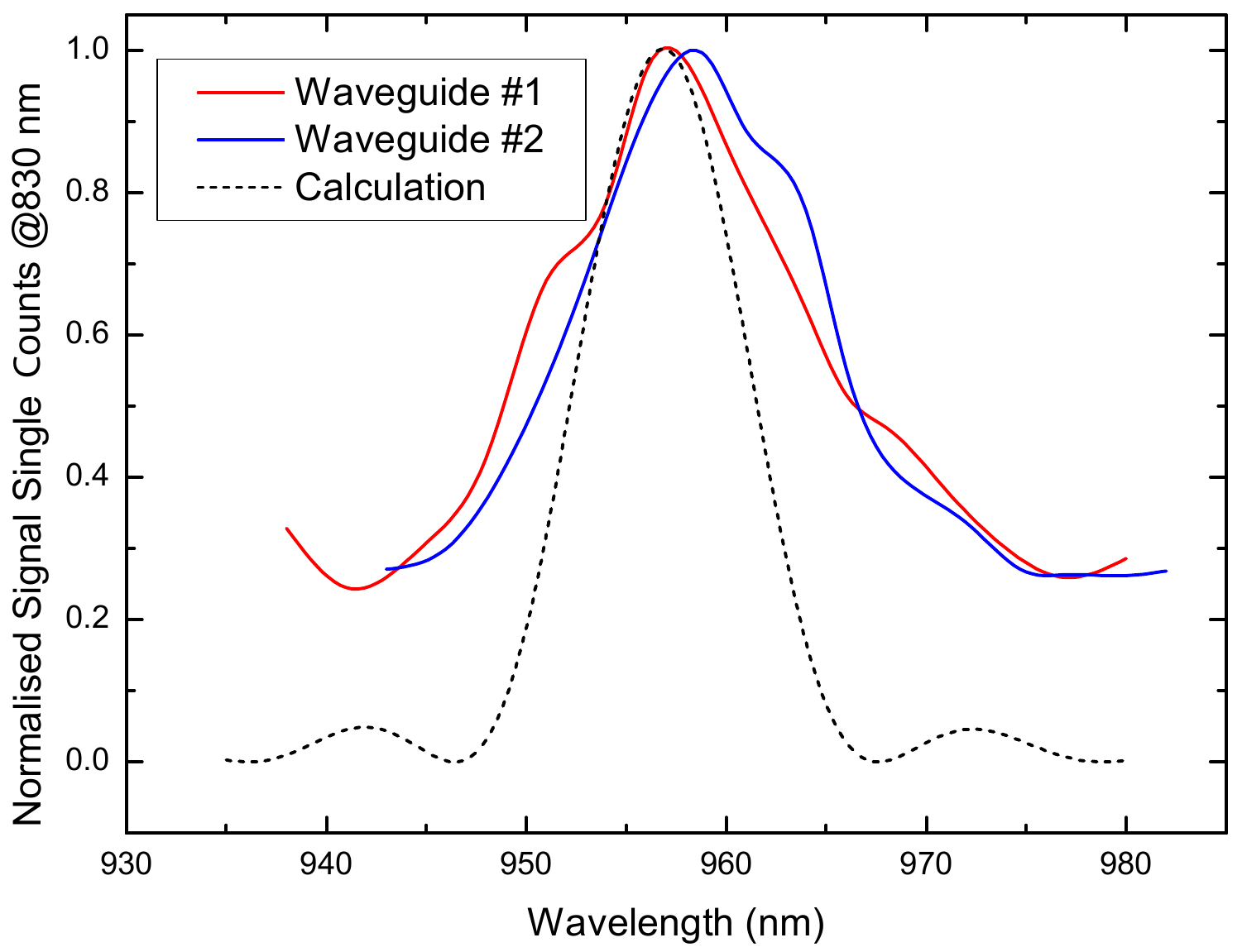}
  \caption{Normalized signal singles counts (at 830~nm) as a function of pump wavelengthin the SFWM regime. The dashed curve shows the numerical calculation of the expected normalised rate $\mathop{\rm sinc}^2(\Delta k L_\text{eff}/2)$ for a birefringence of $1.64 \times 10^{-4}$. The maximum value for waveguide ($\#1$) is 977~$\text{CPS/mW}$, waveguide ($\#2$) is 966~$\text{CPS/mW}$, both at 100 mW pump power.} \label{fig:pmquantumvscalculation}
\end{figure}

\begin{figure*}[!htb]
\centering
 \begin{minipage}{\textwidth}
  \includegraphics[width=0.85\textwidth]{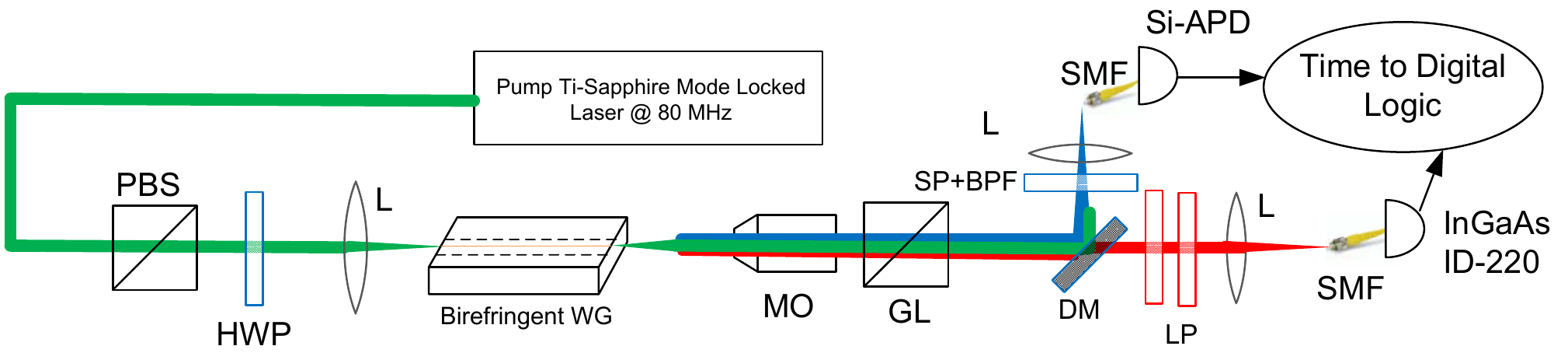}
  \caption{Diagram of the SFWM characterisation setup. MO: Microscope Objective, DM: Dichroic mirror, SMF: Single mode fibre, BPF: Bandpass filter, L: Aspherical lens, LP: Long pass filter, SP: Short pass filter, PBS: Polarisation beam splitter, HWP: Half wave plate, GL: Glan-Laser Calcite Polarizer.  The 50:50 fiber coupler (FC) (not shown) is used to measure the unheralded second order coherence functions.
  }\label{fig:setup}
 \end{minipage}
\end{figure*}

In the SFWM process, the expected photon pair rate is proportional to $\Delta \nu \left( {\gamma P_0 L_\text{eff} } \right)^2 {\mathop{\rm sinc}}^2  \left( {\Delta k L_\text{eff}/2 } \right)$
where $P_0$ is the pump power, and $L_\text{eff}$ is the effective length that is limited by the propagation loss in the FLDW waveguide. The phase-matching term $\Delta k = 2k_p  - k_i  - k_s  + {2 \over 3}\gamma P_0$ is the same form as arises in the classical case, while  $\Delta \nu$ is the effective bandwidth of the photon collection optics, here set  by the 3~nm bandpass filter (BPF) in the signal arm.

Keeping the pump power $P_0$ constant at 100 mW, we let the pump wavelength vary, and monitored the singles count rate for the signal channel at 830~nm. This rate is expected to be modulated by the phase-matching conditions as the pump is tuned. The measured signal counts  are shown as a function of the pump wavelength in Fig.~\ref{fig:pmquantumvscalculation} for two sample waveguides (solid lines). The maximum count rate for the first waveguide~($\text{\#~1}$) occurred at $\lambda_p = 957$~nm. For completeness, a second waveguide~($\text{\#~2}$) was also analysed. In this case the maximum count rate occurred at $\lambda_p = 958$~nm implying excellent reproducibility of the fabrication process, which promises generation of highly indistinguishable heralded photons~\cite{Spring2014}. The dashed curve shows a calculation of the expected $\mathop{\mathrm {sinc}}^2(\Delta k L_\text{eff}/2)$ tuning curve based on the silica Sellmeier dispersion formula~\cite{web}; in this wavelength range, the waveguide contribution to the dispersion is negligible. Based on the calculations in Fig.~\ref{fig:phasematch}, we find that the effective birefringence of the waveguide for this pump band is $\sim$1.64$\times 10^{-4}$, consistent with the classical estimates. For the peak generation at $\lambda_p=957$~nm, the 830~nm signal photon corresponds to an idler photon wavelength of $\lambda_i=$1130~nm.

We then proceeded to characterize the single photon count rates and correlations between the signal and idler photons using the setup shown in Fig.~\ref{fig:setup}.
A narrow BPF (Semrock, centre wavelength $= 830$~nm, full width half maximum (FWHM) $=3$~nm) was used for filtering the signal photons. To suppress the strong pump power, a short pass filter (850~nm cut-off) was also inserted before the detector collection optics. In addition, a Glan-laser calcite polarizer acting as a polarisation filter blocked the vertically polarised pump with an additional attenuation exceeding 30~dB .

\begin{figure}[h]
  \includegraphics[width=\columnwidth]{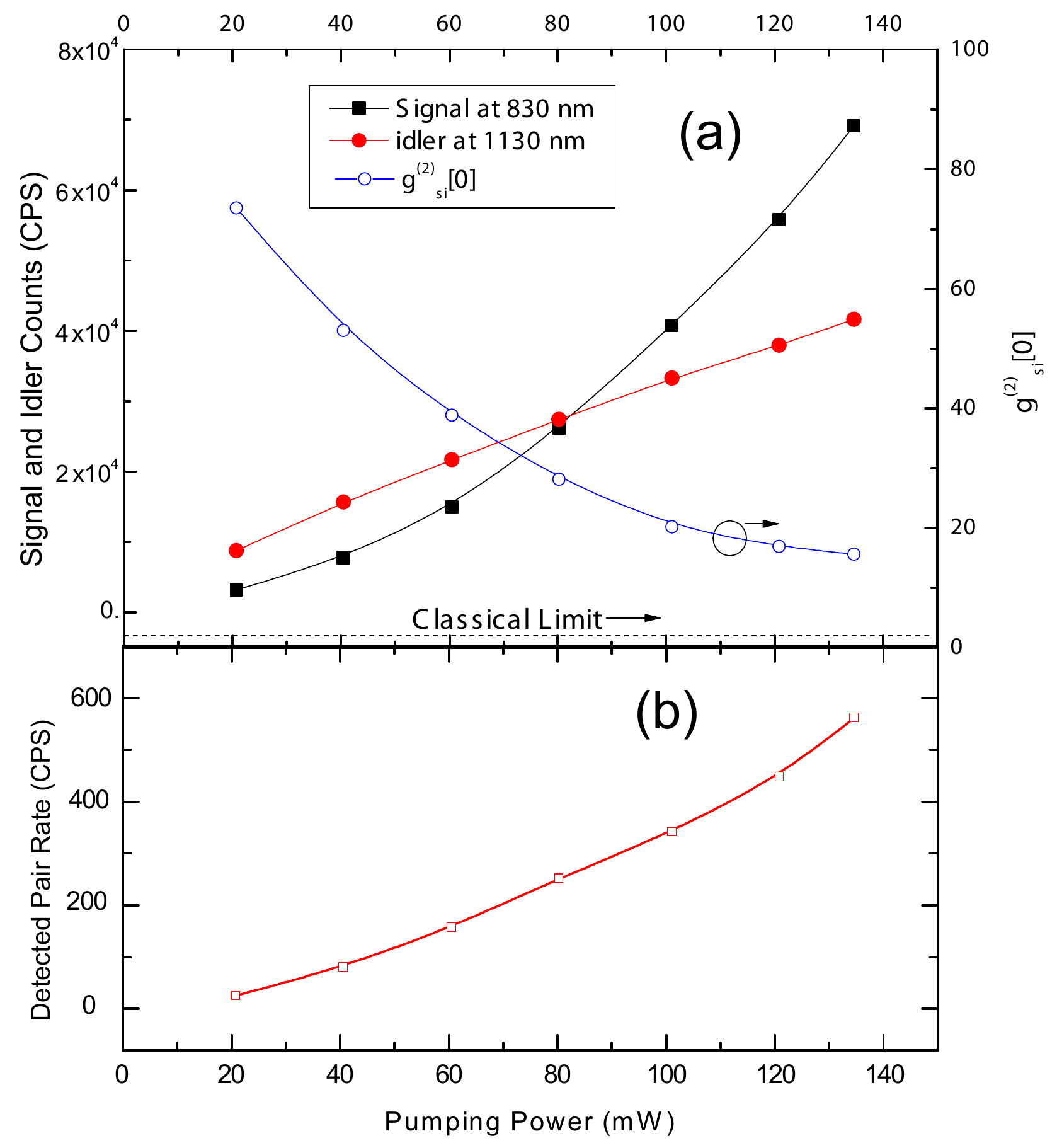}
  \caption{
(a) Left y-axis is the measured counts from the signal (830~nm) and idler (1130~nm) arms. The right y-axis is the cross-second order coherence function. The dash line is the classical limit. The
pump is operated with a 957~nm central wavelength and a 10~nm bandwidth. (b) The pair counting
rate.
  }
  \label{fig:singlesresults}
\end{figure}

The detector for the signal arm is a silicon APD from Excelitas (SPCM-AQ4C), with an estimated efficiency at 830~nm of $\sim$40\%. The detector for the idler arm is an InGaAs infrared free-running single photon detector (id-Quantique id-220, with detection efficiency set to 20$\%$ and 10 $\mu$s dead time).
The signal and idler count rates are shown in Fig.~\ref{fig:singlesresults} (a).
\revision{Note that the much longer dead time of the idler photon detector compared to the
signal detector significantly degrades the effective detection time window when it detects higher rates of photon flux, leading to a reduction in the detection efficiency proportional to the idler photon rate. The detector thus exhibits a sub-quadratic (here approximately linear) dependence on the pump power }
%
%
Fig.~\ref{fig:singlesresults} (b) shows the detected pairs by one silicon APD for signal photons and one InGaAs free-running detector (8\% detector efficiency at 1130 nm) for idler photons. The optical coupling efficiencies are 11\% and 2\% in the idler arm and signal arm, respectively.
\revision{These efficiencies were evaluated by comparing the ratio between the correlated pair counts and the singles counts in each arm excluding both detector inefficiencies. The losses include the contribution from the pair generation setup before the photon detection. The main source of loss is optical coupling from free space to the single mode fiber. The higher loss in the signal arm is associated with reflection loss from the dichroic mirror and the bandpass filtering.}
Taking into account these losses, we estimate the internal pair generation rate to be $\approx 5.05 \times 10^6$ per second at 135~mW pump power.

To confirm the non-classical nature of the nonlinear optical process, we studied the photon correlations through unheralded second order correlation function measurements~\cite{Spring2013}. Non-classical correlations are indicated by the violation of the Cauchy-Schwarz inequality $\left( {g_{si}^{(2)} [0]} \right)^2  \le g_{ss}^{(2)} [0] \cdot g_{ii}^{(2)} [0]$, where $g_{ss}^{(2)} [0]$, $g_{ii}^{(2)} [0]$, $g_{si}^{(2)} [0]$  are the two self and one cross second order correlation functions.

The unheralded $g_{ss}^{(2)} [0]$ in the signal arm is obtained after polarisation filtering and spectral filtering (830/3~nm) to be $1.93 \pm 0.14$ at 135~mW pump power. This indicates that the single photons of the signal arm has a Schmidt mode purity of $\sim 93$~\%. The unheralded $g_{ii}^{(2)} [0]$ is measured to be $1.21\pm 0.50 $ under this condition. Both $g_{ss}^{(2)}[0]$ and $g_{ii}^{(2)} [0]$ were measured by coupling the signal/idler arm photons into a 50:50 fibre coupler individually.
%
%
The specific idler wavelength also limits the choice of suitable bandpass filter.
\revision{If we were able to filter the idler arm to a bandwidth of 3 nm centred at 1130 nm, we would expect a similar Schmidt mode purity of ~93\% as in the signal arm, however this would also reduce the rate of idler photons reaching the detector by around 90\%, as the current unfiltered idler bandwidth is calculated to be approximately 30~nm.}

The cross second order correlation $g_{si}^{(2)} [0]$ at the same pump power is $16.49 \pm 0.43$. Thus the violation of the Cauchy-Schwarz inequality is more than 26 standard deviations at the highest pump power level. Alternatively, the classical limit shown as the dashed line intersecting the right y-axis in Fig.~\ref{fig:singlesresults} (a) was obtained because no unheralded $g_{ss(ii)}^{(2)} [0]$ can exceed 2. From the plot, one can also conclude that the violation of the inequality holds for the entire range of pump powers.

The strong violation of the Cauchy-Schwarz inequality has confirmed that the SFWM result is within the quantum mechanical regime. The experimental SFWM result has demonstrated over 300~nm signal and idler separation. By adding stress tracks on both sides of a very low form birefringence waveguide, the source promises seamless integration with the photonic circuits on the same chip.
The solution in this work provides opportunities not only for realising telecom band idler photons
while keeping signal photons in the efficient APD detection wavelength regime; but also for paving
the critical way for monolithic photon generation and quantum states preparation on FLDW glass
chips.


This research was supported by the ARC Centre of Excellence for Ultrahigh
bandwidth Devices for Optical Systems (project number CE110001018) and was
performed in part at the OptoFab node of the Australian National Fabrication
Facility utilising NCRIS and NSW state government funding.

\end{document}